\documentclass{article}%
\usepackage{amsfonts}
\usepackage{latexsym}
\usepackage{amssymb}
\usepackage{amsmath}
\usepackage{url}
\usepackage{cite}
\usepackage{graphicx}%
\begin{document}

\title{Fly-automata, model-checking and recognizability}
\author{Bruno Courcelle, Ir\`{e}ne Durand\\Bordeaux University, LaBRI, CNRS\\courcell@labri.fr, idurand@labri.fr}
\maketitle

\begin{abstract}
The \emph{Recognizability Theorem} states that if a set of finite graphs is
definable by a monadic second-order (MSO) sentence, then it is recognizable
with respect to the graph algebra upon which the definition of clique-width is
based. Recognizability is an algebraic notion, defined in terms of congruences
that can also be formulated by means of finite automata on the terms that
describe the considered graphs.

This theorem entails that the verification of MSO graph properties, or
equivalently, the model-checking problem for MSO logic over finite binary
relational structures, is fixed-parameter tractable (FPT) for the parameter
consisting of the formula that expresses the property and the clique-width (or
the tree-width) of the input graph or structure. The corresponding algorithms
can be implemented by means of \emph{fly-automata} whose transitions are
computed on the fly and not tabulated.

We review two versions of recognizability, we present fly-automata by means of
examples showing that they can also compute values attached to graphs. We show
that \emph{fly-automata with infinite sets} of states yield a simple proof of
the strong version of the Recognizability Theorem. This proof has not been
published previously.

\end{abstract}


\section*{Introduction}

The \emph{Recognizability Theorem} states that, if a set of finite graphs is
definable by a monadic second-order (MSO) sentence, then it is recognizable
with respect to the graph algebra upon which the definition of
\emph{clique-width} is based. It states a similar result for graphs of bounded
tree-width and the corresponding graph algebra \cite{CouEng}. Recognizability
is defined algebraically in terms of congruences and can also be formulated by
means of finite, or even infinite, automata on the finite terms that describe
the considered graphs. Together with other results (see Chapter 6 of
\cite{CouEng}), this theorem entails that the verification of MSO graph
properties, or equivalently, the model-checking problem for MSO logic over
finite binary relational structures, is fixed-parameter tractable (FPT) for
the parameter consisting of the formula that expresses the property and the
clique-width of the input graph or structure. Tree-width can also be used
instead of clique-width.

\emph{Tree-width} and \emph{clique-width} are graph complexity measures that
serve as parameters in many FPT\ algorithms \cite{CMR, DF, FG} and are based
on hierachical decompositions of graphs.\ These decompositions can be
expressed by terms written with the operation symbols of appropriate graph
algebras \cite{CouEng}. Model-checking algorithms can be based on automata
taking such terms as input. However, the automata associated with
MSO\ formulas, even if they are built for small bounds on tree-width or
clique-width, are in practice much too large to be constructed \cite{FriGro,
Rei}.\ A typical number of states is $2^{2^{10}}$ and lower-bounds match this number.\ 

We overcome this difficulty by using \emph{fly-automata} (\emph{FA})
\cite{BCID12}.\ They are automata whose states are \emph{described} and not
\emph{listed}, and whose transitions are \emph{computed on the fly} and not
\emph{tabulated}.\ When running on a term of size 1000, a deterministic FA
with $2^{2^{10}}$ states computes only 1000\ transitions.\ 

Fly-automata can have infinitely many states.\ For example, a state can
record, among other things, the (unbounded) number of occurrences of a
particular symbol in the input term. FA can thus check some graph properties
that are \emph{not monadic second-order expressible}.\ An example is
\emph{regularity}, the fact that all vertices have the same degree.
Furthermore, an FA equipped with an \emph{output function} that maps the set
of accepting states to an effectively given domain $\mathcal{D}$ can compute a
value, for example the number of $k$-colorings of the given graph $G$, or the
minimum cardinality of one of the $k$ color classes if $G$ is $k$-colorable
(this number measures how close is this graph to be $(k-1)$-colorable). We
have computed with FA the numbers of $k$-colorings for $k=3,4,5$ of some
graphs (cycles, trees, Petersen graph) for which the chromatic polynomial is
known, so that we could test the correctness of the automata (their
correctness can anyway be proved formally).

\bigskip

In this article, we review recognizability, fly-automata and their
applications to the verification of properties or the computation of values
associated with graphs. We present results concerning graphs of bounded
clique-width. Similar results for graphs of bounded tree-width reduce to them
as we will explain.\ In an appendix that can be read as an addendum to
\cite{BCID12}, we explain how the Recognizability Theorem can be proved by
means of fly-automata, in an easier way than in Chapter 5 of \cite{CouEng}.

\bigskip

\bigskip\newpage

\section{Graph algebras, recognizability and automata}

Graphs are finite, undirected, without loops and multiple edges. The extension
to directed graphs, possibly with loops and/or labels is straightforward. A
graph $G$ is identified with the relational structure $\langle V_{G}%
,edg_{G}\rangle$ where $edg_{G}$\ is a binary symmetric relation representing adjacency.

Rather than giving a formal definition of \emph{monadic second-order} (MSO)
logic,\ we present the sentence (i.e., the formula without free variables)
expressing 3-colorability (an NP-complete property).\ It is $\exists
X,Y.Col(X,Y)$ \ where \ $Col(X,Y)$ is the formula

\begin{quote}
$X\cap Y=\emptyset\wedge\forall u,v.\{edg(u,v)\Longrightarrow$

$\qquad\lbrack\lnot(u\in X\wedge v\in X)\wedge\lnot(u\in Y\wedge v\in
Y)\wedge$

$\qquad\qquad\qquad\lnot(u\notin X\cup Y\wedge v\notin X\cup Y)]\}.$
\end{quote}

This formula expresses that $X,Y$ and $V_{G}-(X\cup Y)$ are the three color
classes of a 3-coloring.\ 

\bigskip

\textbf{Definition 1 }: \emph{The graph algebra }$\mathcal{G}$

(a) We will use $\mathbb{N}_{+}$ as a set of labels called \emph{port
labels}.\ A\ \emph{p-graph} is a triple $G=\langle V_{G},edg_{G},\pi
_{G}\rangle$ where $\pi_{G}$ is a mapping : $V_{G}\rightarrow\mathbb{N}_{+}%
$.\ If $\pi_{G}(x)=a,$ we say that $x$ is an $a$-\emph{port}. The set $\pi
(G)$\ of port labels of $G$ is its \emph{type}. By using a default label, say
1, we make every nonempty graph into a p-graph of type $\{1\}$.

(b) For each $k\in\mathbb{N}_{+}$, we define a finite set $F_{k}$ of
\emph{operations} on p-graphs of type included in $C=\{1,...,k\}$ that
consists of :

\begin{quote}
$\bullet$ the binary symbol $\oplus$ denotes the union\ of two \emph{disjoint} p-graphs,

$\bullet$ the unary symbol $relab_{a\rightarrow b}$ denotes the
\emph{relabelling} that changes every port label $a$ into $b$ (where $a,b\in
C$),

$\bullet$ the unary symbol $add_{a,b}$, for $a<b$, $a,b\in C,$ denotes the
\emph{edge-addition} that adds an edge between every $a$-port and every
$b$-port (unless there is already an edge between them; our graphs have no
multiple edges),

$\bullet$ for each $a\in C,$ the nullary symbol $\mathbf{a}$ denotes an
isolated $a$-port.
\end{quote}

(c) Every term $t$ in $T(F_{k})$\ (the set of finite terms written with
$F_{k}$) is called a $k$-\emph{expression}.\ Its \emph{value} is a p-graph,
$val(t)$, that we now define. We denote by $Pos(t)$ the set of
\emph{positions} of $t$: they are the nodes of the syntactic tree of $t$ and
the occurrences of symbols.\ For each $u\in Pos(t)$, we define a p-graph
$val(t)/u$, whose vertex set is the set of leaves of $t$ below $u$. The
definition of $val(t)/u$ is, for a fixed $t$, by bottom-up induction on $u$ :

\begin{quote}
$\bullet$ if $u$ is an occurrence of $\mathbf{a}$, then $val(t)/u$ has vertex
$u$ as an $a$-port and no edge,

$\bullet$ if $u$ is an occurrence of $\oplus$ with sons $u_{1}$ and $u_{2}$, then

$val(t)/u:=$ $val(t)/u_{1}\oplus val(t)/u_{2}$, (note that $val(t)/u_{1}$ and
$val(t)/u_{2}$ are disjoint),

$\bullet$ if $u$ is an occurrence of\ $relab_{a\rightarrow b}$ with son
$u_{1},$ then

$val(t)/u:=relab_{a\rightarrow b}(val(t)/u_{1}),$

$\bullet$ if $u$ is an occurrence of $add_{a,b}$ with son $u_{1},$ then

$val(t)/u:=add_{a,b}(val(t)/u_{1}).$
\end{quote}

Finally, $val(t):=val(t)/root_{t}$. Note that its vertex set is the set of all
leaves (occurrences of nullary symbols). For an example, let

\begin{quote}
$t=\mathit{add}_{b,c}^{1}(\mathit{add}_{a,b}^{2}(\mathbf{a}^{3}\oplus
^{4}\mathbf{b}^{5})\oplus^{6}\mathit{relab}_{b\rightarrow c}^{7}%
(\mathit{add}_{a,b}^{8}(\mathbf{a}^{9}\oplus^{10}\mathbf{b}^{11})))$
\end{quote}

where the superscripts 1 to 11 number the positions of $t$. The\ p-graph
$val(t)$ is$\ 3_{a}-5_{b}-11_{c}-9_{a}$ where the subscripts $a,b,c$ indicate
the port labels. (For clarity, port labels are letters in examples). If $u=2$
and $w=8$, then $t/u=t/w=\mathit{add}_{a,b}(\mathbf{a}\oplus\mathbf{b})$,
however, $val(t)/u$ is the p-graph\ $3_{a}-5_{b}$\ and $val(t)/w$ is
$9_{a}-11_{b}$, isomorphic to $val(t)/u$.

(d) The \emph{clique-width} of a graph $G$, denoted by $cwd(G),$ is the least
integer $k$ such that $G$ is isomorphic to $val(t)$ for some $t$ in $T(F_{k}%
)$. We denote by $\mathcal{G}_{k}$ the set $val(T(F_{k}))$\ of p-graphs that
are the value of a term over $F_{k}.$\ We let $F$ be the union of the sets
$F_{k}$, and $\mathcal{G}$ be the union of the sets $\mathcal{G}_{k}$. Every
p-graph is isomorphic to a graph in $\mathcal{G}$, hence, is defined by some
term hence, has a well-defined clique-width.

(e) An \emph{F-congruence} is an equivalence relation $\approx$ on p-graphs
such that :

\begin{quote}
$\bullet$ two isomorphic p-graphs are equivalent, and

$\bullet$ if $G\approx G^{\prime}$ and $H\approx H^{\prime}$, then $\pi
(G)=\pi(G^{\prime}),$ $add_{a,b}(G)\approx add_{a,b}(G^{\prime}),$
$relab_{a\rightarrow b}(G)\approx relab_{a\rightarrow b}(G^{\prime})$ and
$G\oplus H\approx G^{\prime}\oplus H^{\prime}.$
\end{quote}

(f) A set of graphs $L$ is \emph{recognizable} if it is a (possibly infinite)
union of classes of an $F$-congruence that has finitely many classes of each
finite type $C\subseteq\mathbb{N}_{+}$.

\bigskip

\textbf{Definition 2}: \emph{Fly-automata.}

(a) Let $H$ be a finite or countable, effectively given, signature with arity
mapping denoted by $\rho$. A \emph{fly-automaton} \emph{over }$H$ (in short,
an FA\emph{\ over} $H$)\footnote{A fly-automaton is an automaton on finite
terms whose components are finite or countably infinite and effectively given,
and that has finitely many runs on each term.} is a $4$-tuple $\mathcal{A}%
=\langle H,Q_{\mathcal{A}},\delta_{\mathcal{A}},\mathit{Acc}_{\mathcal{A}%
}\rangle$ such that $Q_{\mathcal{A}} $ is the finite or countable, effectively
given set of \emph{states}, $\mathit{Acc}_{\mathcal{A}}$ is the set of
\emph{accepting states,} a decidable subset of $Q_{\mathcal{A}}$, and
$\delta_{\mathcal{A}}$ is a computable function that defines the
\emph{transition rules}: for each tuple $(f,q_{1},\dots,q_{m})$ such that
$q_{1},\dots,q_{m}\in Q_{\mathcal{A}}$, $f\in H$, $\rho(f)=m\geq0$,
$\delta_{\mathcal{A}}(f,q_{1},\dots,q_{m})$ is a finite set of states.\ We
write $f[q_{1},\dots,q_{m}]\rightarrow q$ (and $f\rightarrow q$ if $f$ is
nullary) to mean that $q\in\delta_{\mathcal{A}}(f,q_{1},\dots,q_{m})$. We say
that $\mathcal{A}$ is \emph{finite} if $F$ and $Q_{\mathcal{A}}$ are
finite.\ Even in this case, it is interesting to have these sets
\emph{specified} rather than listed because this allows to implement finite
automata with huge sets of states \cite{BCID12, BCID13, BCID13a}.

(b) A \emph{run }of $\mathcal{A}$ on a term $t\in T(H)$ is a mapping
$r:Pos(t)\rightarrow Q_{\mathcal{A}}$ such that:

\begin{quote}
if $u\in Pos(t)$ is an occurrence of $f$ with sequence of sons $u_{1}%
,...,u_{m}$, then $r(u)\in\delta_{\mathcal{A}}(f,r(u_{1}),...,r(u_{m})).$
\end{quote}

A run $r$ is \emph{accepting} if $r(root_{t})\in Acc_{\mathcal{A}}.$ A term
$t$ is \emph{accepted} (or \emph{recognized}) by $\mathcal{A}$ if it has an
accepting run. We denote by $L(\mathcal{A})$ the set of terms accepted by
$\mathcal{A}$. A \emph{deterministic} FA $\mathcal{A}$ (by "deterministic" we
mean "deterministic and complete") has a unique run on each term $t$ and
$q_{\mathcal{A}}(t)$\ is the state reached at the root of $t$.\ The mapping
$q_{\mathcal{A}}$\ is computable and the membership in $L(\mathcal{A})$ of a
term $t\in T(H)$ is decidable.

(c) Every FA $\mathcal{A}$ that is not deterministic can be
\emph{determinized} by an easy extension of the usual construction, see
\cite{BCID12}; it is important that the sets $\delta_{\mathcal{A}}(f,q_{1},$
$\dots,q_{m})$\ be finite.

(d) A deterministic FA over $H$ \emph{with output function }is a $4$-tuple
$\mathcal{A}=\langle H,Q_{\mathcal{A}},$ \ $\delta_{\mathcal{A}}%
,\mathit{Out}_{\mathcal{A}}\rangle$\ that is a deterministic FA where
$Acc_{\mathcal{A}}$ is replaced by a total and computable \emph{output
function} $Out_{\mathcal{A}}$: $Q_{\mathcal{A}}\rightarrow\mathcal{D}$ such
that $\mathcal{D}$ is an effectively given domain. The \emph{function computed
by} $\mathcal{A}$ is $Comp(\mathcal{A}):T(H)\rightarrow\mathcal{D}$ such that
$Comp(\mathcal{A})(t):=Out_{\mathcal{A}}(\mathit{q}_{\mathcal{A}}(t))$.\ 

\bigskip

\emph{Example 1}: \emph{The number of accepting runs of an automaton.}

Let $\mathcal{A}=\langle H,Q_{\mathcal{A}},\delta_{\mathcal{A}},\mathit{Acc}%
_{\mathcal{A}}\rangle$ be a nondeterminisic FA. We construct a deterministic
FA $\mathcal{B}$\ that computes the number of accepting runs of $\mathcal{A}$
on any term in $T(H)$. The set of states is the set of finite subsets of
$Q_{\mathcal{A}}\times\mathbb{N}_{+}.$ The transitions are defined so that
$\mathcal{B}$ reaches state $\alpha$ at the root of $t\in T(H)$ if and only if
$\alpha$ is the finite set of pairs $(q,n)\in Q_{\mathcal{A}}\times
\mathbb{N}_{+}$ such that $n$ is the number of runs of $\mathcal{A}$ that
reach state $q$ at the root. This number is finite and $\alpha$\ can be seen
as a partial function : $Q_{\mathcal{A}}\rightarrow\mathbb{N}_{+}$ having a
finite domain. For a symbol $f$ of arity 2, $\mathcal{B}$\ has the transition
: $f[\alpha,\beta]\rightarrow\gamma$ where $\gamma$\ is the set of pairs
$(q,n)$ such that $q\in Q_{\emph{A}}$ and $n$ is the sum of the integers
$n_{p}\times n_{r}$ such that $(p,n_{p})\in\alpha$ and $(r,n_{r})\in\beta.$
\ The transitions for other symbols are defined similarly.\ The function
$\mathit{Out}_{\mathcal{A}}$ maps a state $\alpha$\ to the sum of the integers
$n$ such that $(q,n)\in\alpha\cap(Acc_{\mathcal{A}}\times\mathbb{N}%
_{+}).\square$

\bigskip

\emph{Example 2}: \emph{A fly-automaton\ that checks 3-colorability.}

In order to construct an FA that accepts the terms $t\in T(F)$ such that the
graph $val(t)$ is 3-colorable, we first construct an FA $\mathcal{A}$ for the
property $Col(X,Y)$, taking two sets of vertices $X$ and $Y$ as arguments. For
this purpose, we transform the signature $F$ into $F^{(2)}$ by replacing each
nullary symbol $\mathbf{a}$ by the four nullary symbols $(\mathbf{a},ij)$,
$i,j\in\{0,1\}$. A term $t\in T(F^{(2)})$ defines two things: first, the graph
$val(t^{\prime})$ where $t^{\prime}$ is obtained from $t$ by removing the
Booleans $i,j$ from the nullary symbols and, second, the pair $(X,Y)$ such
that $X$\ is the set of vertices $u$ (leaves of $t$) that are occurrences of
$(\mathbf{a},1j)$ for some $\mathbf{a}$ and $j$, and $Y$\ is the set of those
that are occurrences of $(\mathbf{a},i1)$ for some $\mathbf{a}$ and $i$. The
set of terms $t\in T(F^{(2)})$ such that $Col(X,Y)$ holds in $val(t^{\prime})$
is defined by a deterministic FA $\mathcal{A}$ than we now specify. The
coloring defined by $X,Y$ assigns colors 1,2,3 to the vertices respectively in
$X$, $Y$ and $V_{G}-(X\cup Y)$. Its \emph{type} is the set of pairs $(a,i)$
such that $val(t^{\prime})$ has an $a$-port of color $i$.

We now describe the meaning of the states of $\mathcal{A}$.\ If $u\in Pos(t)$
then $V_{u}$ is the set of vertices of $val(t^{\prime})/u$, i.e., of leaves
below $u$. At position $u$ of $t,$\ the automaton $\mathcal{A}$ reaches the
state $Error$ if and only if $X\cap Y\cap V_{u}\neq\emptyset$ or
$val(t^{\prime})/u$ has\ an edge between two vertices, either both in $X\cap
V_{u}$, or both in $Y\cap V_{u}$, or both in $V_{u}-(X\cup Y)$, hence of same
color, respectively 1,2 or 3; otherwise, $X\cap V_{u}$ and $Y\cap V_{u} $
define a 3-coloring of $val(t^{\prime})/u$ and $\mathcal{A}$ reaches state
$\alpha\subseteq C\times\{1,2,3\}$ where $\alpha$ is the type of this
coloring. All states except $Error$ are accepting.\ Here are the transitions
of $\mathcal{A}$ :

\begin{quote}
$(\mathbf{a},00)\rightarrow\{(a,3)\},(\mathbf{a},10)\rightarrow
\{(a,1)\},(\mathbf{a},01)\rightarrow\{(a,2)\},$

$(\mathbf{a},11)\rightarrow Error.$
\end{quote}

For $\alpha,\beta\subseteq C\times\{1,2,3\}$, $\mathcal{A}$ has transitions :

\begin{quote}
$\oplus\lbrack\alpha,\beta]\rightarrow\alpha\cup\beta,$

$add_{a,b}[\alpha]\rightarrow Error,$ if $(a,i)$ and $(b,i)$ belong to
$\alpha$ for some $i=1,2,3$,

$add_{a,b}[\alpha]\rightarrow\alpha,$ otherwise,

$relab_{a\rightarrow b}[\alpha]\rightarrow\beta,$ obtained by replacing $a $
by $b$ in each pair of $\alpha$.
\end{quote}

Its other transitions are $\oplus\lbrack\alpha,\beta]\rightarrow Error$ if
$\alpha$ or\ $\beta$ is $Error$, $add_{a,b}[Error]\rightarrow Error$ and
$relab_{a\rightarrow b}[Error]\rightarrow Error$.

This FA\ checks $Col(X,Y)$. To check, $\exists X,Y.Col(X,Y),$ we build a
\emph{nondeterministic} FA $\mathcal{B}$\ by deleting the state $Error$ and
the rules containing $Error,$ and by replacing the first three rules of
$\mathcal{A}$\ by $\mathbf{a}\rightarrow\{(a,3)\},\mathbf{a}\rightarrow
\{(a,1)\},\mathbf{a}\rightarrow\{(a,2)\}.$ All states are accepting but on
some terms, no run of $\mathcal{B}$\ can reach the root, and these terms are
rejected. Furthermore, the construction of Example 1 shows how to make
$\mathcal{B}$ into a deterministic FA that computes the number of accepting
runs of $\mathcal{B}$ on a term $t$, hence of 3-colorings of the graph
$val(t)$ because its colorings are in bijection with the accepting runs of
$\mathcal{B}$ on $t$.$\ \square$

\bigskip

\emph{Example 3: Minimal use of one color.}

Continuing Example 2, we want to compute the minimal cardinality of a set $X$
such that $Col(X,Y)$ holds for some set $Y$.\ This cardinality is $\infty$\ if
the considered graph is not 3-colorable.\ It is 0 if it is 2-colorable. We
build from $\mathcal{A}$ a deterministic FA $\mathcal{A}^{\prime}$ over
$F^{(2)}$ \ whose states are $Error$ and the pairs $(\alpha,m)\in
\mathcal{P}(C\times\{1,2,3\})\times\mathbb{N}.$ ($\mathcal{P}(X)$ denotes the
powerset of a set $X$.) The meanings of these states are as for $\mathcal{A}$
except that $m$ in $(\alpha,m)$\ is the cardinality of $X\cap V_{u}$. Some
rules of $\mathcal{A}^{\prime}$ are:

\begin{quote}
$(\mathbf{a},00)\rightarrow(\{(a,3)\},0)$,

$(\mathbf{a},10)\rightarrow(\{(a,1)\},1)$,

$(\mathbf{a},01)\rightarrow(\{(a,2)\},0)$

and $\oplus\lbrack(\alpha,m),(\beta,p)]\rightarrow(\alpha\cup\beta,m+p).$
\end{quote}

We make $\mathcal{A}^{\prime}$ nondeterministic as in Example 2 and we now
detail the deterministic FA\ $\mathcal{C}$\ with output function intended to
compute the minimal cardinality of $X$ such that $Col(X,Y)$ holds for some set
$Y$.

Its states are finite sets of pairs $(\alpha,m)\in\mathcal{P}(C\times
\{1,2,3\})\times\mathbb{N}.$ At the root of a term $t\in T(F)$, the FA
$\mathcal{C}$\ reaches a set $\sigma\subseteq\mathcal{P}(C\times
\{1,2,3\})\times\mathbb{N}$ such that :

\begin{quote}
for each $\alpha\in\mathcal{P}(C\times\{1,2,3\})$ and $m\in\mathbb{N}$, the
pair $(\alpha,m)$ is in $\sigma$ if and only if:

$\alpha$ is the type of a 3-coloring defined by a pair $(X,Y)$,

and $m$ is the minimal cardinality of a set $X$ in such a pair.
\end{quote}

Note that $m$ is uniquely defined from $\alpha$. A state can be defined as a
partial function : $\mathcal{P}(C\times\{1,2,3\})\rightarrow\mathbb{N}.$

The case $\sigma=\emptyset$ corresponds to a graph that is not 3-colorable,
hence, $\emptyset$ plays the role of an $Error$ state.

The transitions of $\mathcal{C}$ are as follows:

\begin{quote}
$\mathbf{a}\rightarrow\{(\{(a,3)\},0),(\{(a,1)\},1),(\{(a,2)\},0)\},$

$\oplus\lbrack\sigma,\sigma^{\prime}]\rightarrow\sigma"$ where $(\gamma
,m)\in\sigma"$ if and only if $m$ is the minimum number $n+n^{\prime}$ such
that $(\alpha,n)\in\sigma$, $(\beta,n^{\prime})\in\sigma^{\prime}$ and
$\alpha\cup\beta=\gamma,$

$add_{a,b}[\sigma]=\sigma^{\prime}$ where $\sigma^{\prime}$ is obtained from
$\sigma$ by removing the pairs $(\alpha,m)$ such that $\alpha$ contains
$(a,i)$ and $(b,i)$ for some $i=1,2,3$,

$relab_{a\rightarrow b}[\sigma]=\sigma^{\prime}$ where $\sigma^{\prime}$ is
obtained by replacing every pair $(a,i)$ occurring in the first component of
any $(\alpha,m)\in\sigma$ by $(b,i).$
\end{quote}

The output function associates with $\sigma$ the minimal $m$ such that
$(\alpha,m)\in\sigma$ for some $\alpha$. If $\sigma=\emptyset$ the output
value is $\infty$ because the graph $val(t)$ is not 3-colorable.

\bigskip

\emph{Remark}: To compute the desired value, we could also use the
determinized automaton of $\mathcal{A}^{\prime}$ with an appropriate output
function.\ Its states encode, for each $\alpha$, the set of cardinalities
$\left\vert X\right\vert $ such that $\alpha$ is the type of a 3-coloring
defined by a pair $(X,Y)$, instead of just the minimal cardinality of such a
set.\ This way, the computation would take more space and more time. $\square$

\bigskip

The constructions of these three examples are particular cases of systematic
and more complex constructions presented in \cite{BCID12, BCID13, BCID13a}.

\bigskip

\section{Two recognizability theorems}

Two theorems relate MSO logic and recognizability.

\bigskip

\textbf{Recognizability Theorem }: The set of graphs that satisfy an MSO
sentence $\varphi$ is $F$-recognizable.

\bigskip

\textbf{Weak Recognizability Theorem }: For every MSO sentence $\varphi$, for
every $k$, the set of graphs in $\mathcal{G}_{k}$ that satisfy $\varphi$ is
$F_{k}$-recognizable.

\bigskip

\emph{About proofs:} The Recognizability Theorem is Theorem 5.68 of
\cite{CouEng}. Its proof shows that the equivalence relation defined by the
fact that two p-graphs have the same type and satisfy the same MSO sentences
of quantifier-height at most that of $\varphi$ satisfies the conditions of
Definition 1(f). The\ Weak Recognizability Theorem follows from the former
one.\ It can also be proved directely by constructing, for each $\varphi$ and
$k$, a finite automaton $\mathcal{A}(\varphi,k)$ (Theorem 6.35 of
\cite{CouEng}).\ One can also construct for each $\varphi$ a single FA
$\mathcal{A}(\varphi)$\ over $F$ that can be seen as the union of the automata
$\mathcal{A}(\varphi,k)$ (\cite{BCID12}). This construction has been
implemented (see below). The proof of the strong theorem in Chapter 5 of
\cite{CouEng} does not provide any usable automaton. As explained in Section
6.4.6 of \cite{CouEng}, the Recognizability Theorem is not a corollary of its
weak form. However, a careful analysis of $\mathcal{A}(\varphi)$\ yields a
simple proof of the Recognizability Theorem as we show in the Appendix.

\section{Other uses of fly-automata}

\emph{Counting and optimizing automata}

Let $P(X_{1},...,X_{s})$ be an MSO property of vertex sets $X_{1},...,X_{s}%
$.\ We denote $(X_{1},...,X_{s})$ by\ $\overline{X}$ and $t\models
P(\overline{X})$ means that $\overline{X}$\ satisfies $P$ in the graph
$val(t)$ defined by a term $t$.\ We are interested, not only to check the
validity of $\exists\overline{X}.P(\overline{X})$, but also to compute from a
term $t$ the following values:

\begin{quote}
$\#\overline{X}.P(\overline{X}),$ defined as the number of assignments
$\overline{X}$ such that

$t\models P(\overline{X}),$

$\mathrm{Sp}\overline{X}.P(\overline{X})$, the \emph{spectrum} of
$P(\overline{X})$, defined as the set of tuples of the form$\ (|X_{1}%
|,\ldots,|X_{s}|)$ such that $t\models P(\overline{X}),$

$\mathrm{MSp}\overline{X}.P(\overline{X})$, the \emph{multispectrum} of
$P(\overline{X})$, defined as the multiset of tuples $(|X_{1}|,\ldots
,|X_{s}|)$ such that $t\models P(\overline{X})$,

the number $\min\{\left\vert Y\right\vert \mid\exists\overline{X}%
.P(Y,\overline{X})\}.$
\end{quote}

These computations can be done by FA \cite{BCID13, BCID13a}.\ We obtain in
this way fixed-parameter tractable (FPT)\ or XP\ algorithms (see \cite{DF, FG}
for the theory of fixed-parameter tractability). A particular case of the
construction for $\#\overline{X}.P(\overline{X})$ is based on Example 1.\ (In
general, the number $\#\overline{X}.P(\overline{X})$ may be larger than the
number of accepting runs of the nondeterministic automaton that checks
$\exists\overline{X}.P(\overline{X})$).

\bigskip

\emph{Beyond MS\ logic}

The property$\ $that the considered graph is the union of two disjoint regular
graphs with possibly some edges between these two subgraphs is not
MSO\ expressible but can be checked by an FA. An FA can also compute the
minimal number of edges between $X$ and $V_{G}-X$\ such that $G[X]$ and
$G[V_{G}-X]$ are connected, when such a set $X$ exists.

\bigskip

\emph{Edge set quantifications and tree-width.}

The incidence graph $Inc(G)$ of a graph $G$ (that can have multiple edges) is
a bipartite graph whose vertex set is $V_{G}\cup E_{G}$\ where $E_{G}$\ is the
set of edges of $G$, and $Inc(G)$ has an edge between $x\in V_{G}$ and $e\in
E_{G}$ if and only if $x$ is an end of $e$. An MSO sentence evaluated in
$Inc(G)$ is thus able to use quantifications on edges and sets of edges.\ The
graph properties expressed by such sentences are said to be MSO$_{2}$
expressible. That $G$ is Hamiltonian can be expressed by "there exists a set
of edges that forms a Hamiltonian cycle", hence is MSO$_{2}$ expressible,
whereas this property is not MSO\ expressible.\ If $G$ has tree-width $k$,
then $Inc(G)$ has clique-width at most $3.2^{k}$ (see \cite{CouEng},
Proposition 2.114)\ and even at most $k+3$ by a recent unpublished result due
to T.\ Bouvier (LaBRI).\ Furthermore, a term defining $Inc(G)$ that witnesses
the latter upperbound can be obtained in linear time from a tree-decomposition
of $G$ of width $k$. It follows that FA\ can be used to verify MSO$_{2}$
expressible properties of graphs of bounded tree-width. Counting and
optimizing functions based on such properties can also be computed by FA.
Another approach is in \cite{Cou12}.

The two recognizability theorems have versions for MSO$_{2}$ expressible
properties of graphs of bounded tree-width (see \cite{CouEng}, Theorems 5.68
and 5.69).

\bigskip

\section{Experimental results and open problems}

These constructions have been implemented and tested\footnote{AUTOGRAPH\ is
written in Steele Bank Common Lisp and computations have been done on a Mac
Book Pro (Mac OS X 10.9.4 with processor Intel Core 2 Duo, 2.53 GHz and memory
of 4 GB, 1067 MHz DDR3).} \cite{BCID12, BCID13, BCID13a}. We have computed the
number of optimal colorings of some graphs of clique-width at most 8 for which
the chromatic polynomial is known, which allowed us to verify the correctness
of the automaton. We could verify in, respectively, 35 and 105\ minutes that
the 20$\times$20 and the 6$\times$60\ grids are 3-colorable. In 29 minutes, we
could verify that the McGee graph (24 vertices) given by a term over $F_{8}$
is acyclically 3-colorable.

\bigskip

A different model-checking method based on games is presented in \cite{KLR}.
It gives a proof of the Weak Recognizability Theorem for graphs of bounded
tree-width and has also been implemented and tested.

\bigskip

The \emph{parsing problem} for graphs of clique-width at most $k$ is
NP-complete (with $k$ in the input) \cite{Fell+}.\ Good heuristics remain to
be developped.

\section*{Appendix}

We explain to the reader familiar with \cite{BCID12} how the Recognizability
Theorem can be proved from the construction, for every MSO sentence $\varphi$,
of an FA $\mathcal{A}(\varphi)$\ over $F$ that recognizes the terms whose
value is a finite model of $\varphi$. This is a new proof of this theorem. We
first review definitions and notation.

\bigskip

MSO\ formulas are written with set variables $X_{1},...,X_{n},...$ (without
first-order variables), the atomic formulas $X_{i}\subseteq X_{j}$,
$Sgl(X_{i})$ (meaning that $X_{i}$ is singleton), $edg(X_{i},X_{j})$ (meaning
that $X_{i}$ and $X_{j}$ are singletons consisting of adjacent vertices),
negation, conjunction and existential quantifications of the form $\exists
X_{n}.\psi$ where $\psi$ has its free variables among $X_{1},...,X_{n}$.

Generalizing the definition of Example 2, we transform $F$ into $F^{(m)}$ (for
$m>0$) by replacing each nullary symbol $\mathbf{a}$ by the nullary symbols
$(\mathbf{a},w)$ for all $w\in\{0,1\}^{m}$. Hence, a term $t\in T(F^{(m)})$
defines the p-graph $val(t^{\prime})$ where $t^{\prime}$ is obtained from $t$
by removing the sequences $w$ from the nullary symbols and the $m$-tuple
$(V_{1},...,V_{m})$ such that $V_{i}$\ is the set of vertices $u$ (leaves of
$t$) that are occurrences of $(\mathbf{a},w)$ for some $\mathbf{a}$ and $w$
with 1 at its $i$-th position.\ We denote\ $val(t^{\prime})$ by $val(t)$ and
$(V_{1},...,V_{m})$ by $\nu(t)$.

If $\varphi$\ is an MSO formula with free variables among $X_{1},...,X_{m}$,
we let $L(\varphi,X_{1},$ $...,X_{m})$\ \ be the set of terms $t\in
T(F^{(m)})$ such that $(val(t),\nu(t))\models\varphi$.

\bigskip

\textbf{Theorem} \cite{BCID12} : Let $\varphi$\ be an MSO formula with free
variables among $X_{1},...,$\ \ $X_{m}$.\ One can construct a fly-automaton
$\mathcal{A}(\varphi,X_{1},...,X_{m})$ over $F^{(m)}$ that recognizes the set
$L(\varphi,X_{1},...,X_{m})$.

\bigskip

We revisit this construction to prove the Recognizability Theorem. For a
finite set $B$\ and an integer $i\geq0$, we define the finite set
$\mathcal{L}_{i}(B)$ as follows:

\begin{quote}
$\mathcal{L}_{0}(B):=B,$

$\mathcal{L}_{i+1}(B):=\mathcal{L}_{i}(B)\cup\mathcal{P}(\mathcal{L}%
_{i}(B))\cup\mathcal{L}_{i}(B)\times\mathcal{L}_{i}(B).$
\end{quote}

\ In order to have a unique notation for the elements of these sets, we write
an element of $\mathcal{P}(\mathcal{L}_{i}(B))$ as $\{w_{1},...,w_{p}\} $ with
the condition that $w_{1}<w_{2}<...<w_{p}$ for some lexicographic order
$<$
on the words denoting the elements of the sets $\mathcal{L}_{n}(B).$

\bigskip

The proof of the previous theorem yields the following more precise statement.

\bigskip

\textbf{Proposition }: Let $\varphi$\ be an MSO formula with free variables
among $X_{1},...,$ \ $X_{m}$.\ One can construct a finite set $B$ disjoint
from $\mathbb{N}_{+}$, an integer $i$ and a deterministic fly-automaton
$\mathcal{A}(\varphi,X_{1},...,X_{m})$ over $F^{(m)}$ that recognizes the set
$L(\varphi,X_{1},...,X_{m})$ and satisfies the following two properties, for
all $t,t^{\prime}\in T(F^{(m)})$:

\begin{quote}
(i) $q_{\mathcal{A}(\varphi,X_{1},...,X_{m})}(t)\in\mathcal{L}_{i}(B\cup
\pi(val(t))),$

(ii) if $(val(t),\nu(t))$ is isomorphic to $(val(t^{\prime}),\nu(t^{\prime
})),$ then $q_{\mathcal{A}(\varphi,X_{1},...,X_{m})}(t)=q_{\mathcal{A}%
(\varphi,X_{1},...,X_{m})}(t^{\prime}).$
\end{quote}

\bigskip

\textbf{Proof }: The proof is by induction on the structure of\ $\varphi$. We
assume that the reader has access to \cite{BCID12}, so we will not detail the automata.

If $\varphi$\ is $X_{i}\subseteq X_{j}$ or $Sgl(X_{i}),$\ then the states of
$\mathcal{A}(\varphi,X_{1},...,X_{m})$ do not use port labels, hence
$q_{\mathcal{A}(\varphi,X_{1},...,X_{m})}(t)\in\mathcal{L}_{0}(B)$ for some
finite set $B$. Properties (i) and (ii) hold.

If $\varphi$\ is $edg(X_{i},X_{j})$, then the states of $\mathcal{A}%
(\varphi,X_{1},...,X_{m})$ are $Error,Ok$, $(a,\underline{1})$, $(a,\underline
{2})$, $(a,b)$ for $a,b\in\mathbb{N}_{+}$, hence belong to $B\cup
\mathbb{N}_{+}\times(\mathbb{N}_{+}\cup B)$ where $B=\{Error,Ok,$
\ $\underline{1},\underline{2}\}$. (To be precise the states $(a,\underline
{1}),(a,\underline{2})$ and $(a,b)$ are written respectively $a(1),a(2)$ and
$ab $ in \cite{BCID12}.) The port labels occurring in the state
$q_{\mathcal{A}(edg(X_{i},X_{j}),X_{1},...,X_{m})}(t)$ are in $\pi(val(t))$:
this is clear from the meanings of the states described in Table 3\ of
\cite{BCID12}.\ So we have $q_{\mathcal{A}(\varphi,X_{1},...,X_{m})}%
(t)\in\mathcal{L}_{1}(B\cup\pi(val(t))).$ The validity of (ii) is also clear
from the same table.

\bigskip

If $\varphi$\ is $\lnot\psi$\ and since we construct deterministic (and
complete) automata, $\mathcal{A}(\varphi,X_{1},...,X_{m})$ and $\mathcal{A}%
(\psi,X_{1},...,X_{m})$ differ only in their accepting states.\ Hence
$\mathcal{A}(\varphi,X_{1},...,X_{m})$ satisfies Properties (i) and (ii) since
$\mathcal{A}(\psi,X_{1},...,X_{m})$ does.

\bigskip

If $\varphi$\ is $\theta\wedge\psi,$\ then $\mathcal{A}(\varphi,X_{1}%
,...,X_{m})$ is the product automaton of $\mathcal{A}(\theta,X_{1},...,$
$X_{m})$\ \ and $\mathcal{A}(\psi,X_{1},...,X_{m})$ (in particular
$\ Q_{\mathcal{A}(\varphi,X_{1},...,X_{m})}=Q_{\mathcal{A}(\theta
,X_{1},...,X_{m})}\times Q_{\mathcal{A}(\psi,X_{1},...,X_{m})}$).\ If $(B,i)$
and $(B^{\prime},j)$ are associated by induction with $\theta$ and $\psi$,
then we can take the pair $(B\cup B^{\prime},1+\max(i,j))$ for $\varphi$,
which gives Property (i). Property (ii) is easy to check.

\bigskip

If $\varphi$\ is $\exists X_{m}.\psi$ where $\psi$ has its free variables
among $X_{1},...,X_{m}$, then $\mathcal{A}(\varphi,X_{1},$ $...,X_{m-1}%
)$\ \ is obtained from $\mathcal{A}(\psi,X_{1},...,X_{m})$ as follows:

\begin{quote}
(1) one builds an FA $\mathcal{B}$\ by replacing in $\mathcal{A}(\psi
,X_{1},...,X_{m})$ all transitions $(\mathbf{a},w0)\rightarrow p$ and
$(\mathbf{a},w1)\rightarrow q$ by $(\mathbf{a},w)\rightarrow p$ and
$(\mathbf{a},w)\rightarrow q$ so that $\mathcal{B}$ is not deterministic;

(2) $\mathcal{A}(\varphi,X_{1},...,X_{m-1})$ is defined as the determinized
automaton of $\mathcal{B}$.
\end{quote}

If $q_{\mathcal{A}(\psi,X_{1},...,X_{m})}(t)\in\mathcal{L}_{i}(B\cup
\pi(val(t))),$ then $q_{\mathcal{A}(\varphi,X_{1},...,X_{m-1})}(t)\in
\mathcal{P}(\mathcal{L}_{i}(B\cup\pi(val(t))))\subseteq\mathcal{L}_{i+1}%
(B\cup\pi(val(t))),$ which proves (i). Property (ii) is easy to check.

\bigskip

It may be necessary to construct $\mathcal{A}(\varphi,X_{1},...,X_{m})$ from
$\mathcal{A}(\varphi,X_{1},...,X_{n})$ where $m>n.$ A typical example is for
$\theta=\varphi\wedge\psi$\ in a case where we have already constructed
$\mathcal{A}(\varphi,X_{1},X_{2})$ and $\mathcal{A}(\psi,X_{1},X_{2},X_{3})$;
we must take the product of $\mathcal{A}(\varphi,X_{1},X_{2},X_{3})$ and
$\mathcal{A}(\psi,X_{1},X_{2},X_{3})$.\ This situation is handled by Lemma 13
and Definition 17(h) of \cite{BCID12}: the automaton $\mathcal{A}%
(\varphi,X_{1},...,X_{m})$ has the same states as $\mathcal{A}(\varphi
,X_{1},...,X_{n})$ and Properties (i) and (ii) are inherited from
$\mathcal{A}(\varphi,X_{1},...,X_{n})$. $\square$

\bigskip

\textbf{Proof of the Recognizability Theorem: \ }Let $\varphi$ be an MSO
sentence and $\mathcal{A}(\varphi),B$\ and $i$ be constructed by the previous proposition.\ 

Let $G$ be a p-graph and $t$ be any term in $T(F)$ that defines it. Then, by
Property (ii), the state $q_{\mathcal{A}(\varphi)}(t)$ depends only on
$G$\ (it is the same for every term $t$ that defines $G$), hence can be
written $q(G)$. By (i), $q(G)\in\mathcal{L}_{i}(B\cup\pi(G)).$

We define an equivalence relation by :

\begin{center}
$G\approx G^{\prime}$ if and only if $\pi(G)=\pi(G^{\prime})$ and
$q(G)=q(G^{\prime})$.
\end{center}

Two isomorphic graphs are equivalent by Property (ii) and two equivalent
graphs have the same type. We prove that\ $\approx$ is a congruence.

Let $G\approx G^{\prime}$ and $H\approx H^{\prime}$.\ Then $\pi(G\oplus
H)=\pi(G^{\prime}\oplus H^{\prime})=\pi(G)\cup\pi(H)$.

Let $t_{G}$ define $G$ and $t_{G^{\prime}}$ define $G^{\prime}$. Then
$q(G)=q_{\mathcal{A}(\varphi)}(t_{G})=q(G^{\prime})=q_{\mathcal{A}(\varphi
)}(t_{G^{\prime}})$ and similarly for $H$ and $H^{\prime}$. We have $q(G\oplus
H)=q_{\mathcal{A}(\varphi)}(t_{G}\oplus t_{H})$ and so, $\oplus\lbrack
q_{\mathcal{A}(\varphi)}(t_{G}),q_{\mathcal{A}(\varphi)}(t_{H})]\rightarrow
_{\mathcal{A}(\varphi)}q(G\oplus H).$\ Similarly, $\oplus\lbrack
q_{\mathcal{A}(\varphi)}(t_{G^{\prime}}),q_{\mathcal{A}(\varphi)}%
(t_{H^{\prime}})]$ $\rightarrow q(G^{\prime}\oplus H^{\prime})\ $and
$q(G^{\prime}\oplus H^{\prime})=q(G\oplus H)$ since $\mathcal{A}(\varphi)$ is
deterministic. Hence, $G\oplus H\approx G^{\prime}\oplus H^{\prime}.$ The
proof is similar for all unary operations.

Since $\mathcal{L}_{i}(B\cup C)$ is finite for $C$ finite, the congruence
$\approx$ has finitely many classes of each finite type $C\subseteq
\mathbb{N}_{+}$.

A p-graph $G$ satisfies $\varphi$\ if and only if $q(G)$ is an accepting state
of $\mathcal{A}(\varphi)$.\ Hence the set of finite models of $\varphi$\ is a
union of classes of $\approx$, hence is recognizable. $\square$

\bigskip

In \cite{BCID12}, we have\ constructed FA for other basic properties than
$X_{i}\subseteq X_{j}$, $Sgl(X_{i})$\ and $edg(X_{i},X_{j})$, and in
particular, for $Card_{p}(X_{1})$ ($X_{1}$\ has $p$ elements),
$Partition(X_{1},...,X_{m})$ ($X_{1},...,X_{m}$ is a partition of the vertex
set), $Path(X_{1},X_{2})$ ($X_{1}$ consists of two vertices linked by a path
having its vertices in $X_{2}$), connectedness and existence of cycles.\ These
FA satisfy Properties (i) and (ii) : the proofs are the same as for
$X_{i}\subseteq X_{j}$, $Sgl(X_{i})$\ and $edg(X_{i},X_{j}).$\ However, the
minimal syntax for MSO\ formula that we use is enough to prove the
Recognizability Theorem.

The construction of FA\ for the properties $Card_{p,q}(X_{1})$ ($X_{1}$\ has
$p$ modulo $q$ elements) yields the proof of the Recognizability Theorem for
\emph{counting monadic second-order}\ logic.\ See \cite{CouEng} for details.

In \cite{BCID12}, we have also constructed "smaller" FA that work correctly on
terms in $T(F)$ satisfying the special condition to be \emph{irredundant} (no
edge is created between two vertices $x$ and $y$ if there exists already one).
For an automaton on irredundant terms, the state reached at some node $u$ of a
term $t$ does not depend only on the graph $val(t)/u$ but also, implicitly, on
the context of $u$ in $t$. These automata are useful for model-checking
because they are smaller than the equivalent general ones and terms can be
preprocessed appropriately, but they may not satisfy Property (ii).\ Hence,
they cannot be used in the above proof of the Recognizability Theorem.\ 
\end{document}